\documentclass[aoas, preprint]{imsart}
\usepackage{amsmath,amssymb,fullpage,fouriernc,color,url,graphicx}
\usepackage[authoryear]{natbib}
\usepackage{amsthm}
\usepackage[colorlinks,citecolor=blue,urlcolor=blue]{hyperref}

\usepackage{algorithm}% http://ctan.org/pkg/algorithms
\usepackage{algpseudocode}% http://ctan.org/pkg/algorithmicx
\algnewcommand\algorithmicinput{\textbf{INPUT:}}
\algnewcommand\INPUT{\item[\algorithmicinput]}
\algnewcommand\algorithmicoutput{\textbf{OUTPUT:}}
\algnewcommand\OUTPUT{\item[\algorithmicoutput]}

\usepackage{color}

\newcommand{\argmin}{\mathop{\mathrm{arg\,min}}}
\newcommand{\argmax}{\mathop{\mathrm{arg\,max}}}

\def\P{\mathbb{P}}

\def\hbeta{\hat\beta}

\def\R{\mathbb{R}}

\def\prox{\mathrm{prox}}

\begin{document}

\begin{frontmatter}

\title{High-dimensional Longitudinal Classification\\with the Multinomial Fused Lasso}
\runtitle{Multinomial Fused Lasso}

\begin{aug}
\author{\fnms{Samrachana}
  \snm{Adhikari}\thanksref{a,t1}\ead[label=e1]{asamrach@andrew.cmu.edu}},
\author{\fnms{Fabrizio}
  \snm{Lecci}\thanksref{a, t1}\ead[label=e2]{lecci@cmu.edu}}, 
\author{\fnms{James T.}
  \snm{Becker}\thanksref{b}\ead[label=e3]{beckerJT@upmc.edu}}, 
\author{\fnms{Brian W.}
  \snm{Junker}\thanksref{a}\ead[label=e4]{brian@stat.cmu.edu}},
\author{\fnms{Lewis H.}
  \snm{Kuller}\thanksref{b}\ead[label=e5]{kullerl@edc.pitt.edu}},
\author{\fnms{Oscar L.}
  \snm{Lopez}\thanksref{b}\ead[label=e6]{lopezOL@upmc.edu}} 
\and
\author{\fnms{Ryan J.}
  \snm{Tibshirani}\thanksref{a}\ead[label=e7]{ryantibs@cmu.edu}}

\thankstext{t1}{These authors contributed equally to this work}
%%\thankstext{t2}{Research supported xxxx.}
%\thankstext{t3}{Research supported xxxx.}
%\thankstext{t4}{Research supported xxxx.}
%\thankstext{t5}{Research supported xxxx.}
%\thankstext{t6}{Research supported xxxx.}

\runauthor{Adhikari et al.}

\affiliation[a]{Carnegie Mellon University}
\affiliation[b]{University of Pittsburgh}

\address{
S. Adhikari\\
F. Lecci\\
B.W. Junker\\
R.J. Tibshirani\\
Department of Statistics\\
Carnegie Mellon University\\
Pittsburgh, PA 15213\\
\printead{e1}\\
\phantom{E-mail:\ }\printead*{e2}\\
\phantom{E-mail:\ }\printead*{e4}\\
\phantom{E-mail:\ }\printead*{e7}
}

\address{
L.H. Kuller\\
Department of Epidemiology\\
University of Pittsburgh\\
Pittsburgh, PA 15213\\
\printead{e5}
}

\address{
J.T. Becker\\
O.L. Lopez\\
Department of Neurology\\
University of Pittsburgh\\
Pittsburgh, PA 15213\\
\printead{e3}\\
\phantom{E-mail:\ }\printead*{e6}
}

\end{aug}

\begin{abstract}
We study regularized estimation in high-dimensional longitudinal
classification problems, using the lasso and fused lasso regularizers.
The constructed coefficient estimates are piecewise constant across
the time dimension in the longitudinal problem, with adaptively
selected change points (break points).  We present an efficient
algorithm for computing such estimates, based on proximal gradient
descent.  We apply our proposed technique to a longitudinal data set
on Alzheimer's disease from the Cardiovascular Health Study Cognition
Study, and use this data set to motivate and demonstrate several
practical considerations such as the selection of tuning parameters,
and the assessment of model stability. 
\end{abstract}

%\begin{keyword}[class=MSC]
%\kwd[Primary ]{60K35}
%\kwd{60K35}
%\kwd[; secondary ]{60K35}
%\end{keyword}

\end{frontmatter}

%%%%%%%%%%%%%%%%%%%%%
\section{Introduction}
%%%%%%%%%%%%%%%%%%%%%
\label{sec::introduction}

In this paper, we study longitudinal classification problems in which
the number of predictors can exceed the number of  observations.  The
setup: we observe $n$ individuals across discrete timepoints
$t=1,\ldots T$. At each timepoint we record $p$ predictor variables
per individual, and an outcome that places each individual into one of
$K$ classes.  The goal is to construct a model that predicts the
outcome of an individual at time $t+\Delta$, given his or her
predictor measurements at time $t$. 
Since we allow for the possibility that $p>n$, regularization must be
employed in order for such a predictive model (e.g., based on maximum
likelihood) to be well-defined.  Borrowing from the extensive
literature on high-dimensional regression, we consider two well-known
regularizers, each of which also has a natural place in
high-dimensional longitudinal analysis for many scientific problems of
interest.  The first is the {\it lasso} regularizer, which encourages
overall sparsity in the active (contributing) predictors at each
timepoint; the second is the {\it fused lasso} regularizer, which
encourages a notion of persistence or contiguity in the sets of active
predictors across timepoints.   

% In each timepoint, we allow for the possibility that 
% We believe that only a small number of predictor variables are important (sparsity).  We also believe that predictors that play an important role at one timepoint are likely to play an important role at neighboring timepoints (persistence).
% Our model will use the important variables recorded at time $t$ to predict the outcome variable at time $t +\Delta$. 
% In Figure \ref{fig:model} we represent our objective in a simplified situation ($K=2$): we build a sequence of logistic regression models for the binary variable $Y$; for each timepoint $t$ we want to find a subset of predictor variables that are relevant for the prediction of $Y^{(t+\Delta)}$, assuming that the variables that are important at time $t$ are likely to be important at some consecutive timepoints.

% \begin{figure}[!h]
%  \centering
%  \includegraphics[scale=0.75]{figs/model2.pdf}
%  \caption{Sparsity and Persistence of the coefficients $\beta_j$ in consectuive Logistic Regression models.}
%  \label{fig:model}
% \end{figure}

Our work is particularly motivated by the analysis of a large data set
provided by the Cardiovascular Health Study Cognition Study (CHS-CS).
Over the past 24 years, the CHS-CS recorded multiple metabolic,
cardiovascular and neuroimaging risk factors for Alzheimer's disease
(AD), as well as detailed cognitive assessments for people of ages 65
to 110 years old \citep{lopez2003prevalence, saxton2004preclinical,
  lopez2007incidence}.  
%{\todo{Do we need an authorization for the CHS data?}}  
As a matter of background, the prevalence of AD increases at an
exponential-like rate beyond the age of 65. 
%\todo{What is the prevalence for AD with 65 year olds?}
After 90 years of age, the incidence of AD increases dramatically,
from 12.7\% per year in the 90-94 age group, to 21.2\% per year in the
95-99 age group, and to 40.7\% per year for those older than 100 years
\citep{evans1989prevalence, fitzpatrick2004incidence,
  corrada2010dementia}.  Later, we examine data from 924 individuals
in the Pittsburgh section of the CHS-CS.   
% The interactions of multiple risk factors and the pathobiological
% cascade of AD determines the likelihood of a clinical expression of
% AD either as dementia or Mild Cognitive Impairment (MCI).  
%Of course, AD is a complex disease and our statistical perspective is
%only a simplification of the scientific problem, but (we hope) an
%interesting perspective nonetheless.   
The objective is to use the data available from subjects at $t$ years
of age to predict the onset of AD at $t+10$ years of age
($\Delta=10$). For each age, the outcome variable assigns an
individual to one of 3 categories: normal, dementia, death.  Refer to
Section \ref{sec::CHS} for our analysis of the CHS-CS data set. 

\subsection{The multinomial fused lasso model}

Given the number of parameters involved in our general longitudinal
setup, it will be helpful to be clear about notation: see Table
\ref{tab:notation}.  Note that the matrix $Y$ stores future outcome
values, i.e., the element $Y_{it}$ records the outcome of the
$i$th individual at time $t+\Delta$, where $\Delta \geq 0$ determines
the time lag of the prediction. In the following, we will generally
use the ``$\cdot$'' symbol to denote partial indexing; examples are
$X_{i\cdot t}$, the vector of $p$ predictors for individual $i$ at
time $t$, and $\beta_{\cdot tk}$, the vector of $p$ multinomial 
coefficients at time $t$ and for class $k$.  Also, Section
\ref{sec::GGD} will introduce an extension of the basic setup in which
the number of individuals can vary across timepoints, with $n_t$
denoting the number of individuals at each timepoint $t=1,\ldots T$. 

\begin{table}[h!]
\centering
\begin{tabular}{| c | c |}
\hline
\textbf{Parameter} & \textbf{Meaning}  \\ 
\hline \hline
$i=1, \ldots n$ & index for individuals \\ \hline
$j=1, \ldots p$ & index for predictors \\ \hline 
$t=1,\ldots T$ & index for timepoints \\ \hline
$k=1, \ldots K$ & index for outcomes  \\ \hline \hline
$Y$ & $n \times T$ matrix of (future) outcomes \\ \hline
$X$ & $n \times p \times T$ array of predictors\\ \hline
$\beta_0$ & $T \times (K-1)$ matrix of intercepts  \\ \hline
$\beta$ & $p \times T \times (K-1)$ array of coefficients  \\ \hline
\end{tabular}
\caption{Notation used throughout the paper.}
\label{tab:notation}
\end{table}

At each timepoint $t=1,\ldots T$, we use a separate multinomial logit 
model for the outcome at time $t+\Delta$: 
%We want to estimate the coefficients stored in the array $\beta$
%using $T$ independent logistic regression models.  
%At each time $t$ the logistic regression model for $K$ classes is 
%expressed through $K-1$ equations (see
%%\cite{hastie01statisticallearning}): 
\begin{equation}
\begin{aligned}
\label{eq::multinomial}
\log \frac{\P(Y_{it}=1 | X_{i\cdot t} = x)}{\P(Y_{it}=K | X_{i\cdot t}
  = x)}&= \beta_{0t1} + \beta_{\cdot t1}^T x \\ 
\log \frac{\P(Y_{it}=2 | X_{i\cdot t} = x)}{\P(Y_{it}=K | X_{i\cdot t}
  = x)}&= \beta_{0t2} + \beta_{\cdot t2}^T x \\ 
& \vdots \\
\log \frac{\P(Y_{it}=K-1 | X_{i\cdot t} = x)}{\P(Y_{it}=K | X_{i\cdot
    t} = x)}&= \beta_{0t(K-1)} + \beta_{\cdot t(K-1)}^T x. 
\end{aligned}
\end{equation}
The coefficients are determined by maximizing a penalized log
likelihood criterion, 
\begin{equation}
\label{eq::fusedmodel}
(\hbeta_0, \hbeta) \in \argmax_{\beta_0,\beta} \; 
\ell(\beta_0,\beta) - \lambda_1 P_1(\beta) - \lambda_2 P_2(\beta), 
\end{equation}
where $\ell(\beta_0,\beta)$ is the multinomial log likelihood,
\begin{equation*}
\ell(\beta_0,\beta) = \sum_{t=1}^T \sum_{i=1}^n \P(Y_{it}|X_{i\cdot
  t}), 
\end{equation*}
$P_1$ is the lasso penalty \citep{lasso}, 
\begin{equation*}
P_1(\beta) =\sum_{j=1}^p  \sum_{t=1}^T \sum_{k=1}^{K-1}
|\beta_{jtk}|. 
\end{equation*}
and $P_2$ is a version of the fused lasso penalty \citep{fuse} applied
across timepoints, 
\begin{equation*}
P_2(\beta) = \sum_{j=1}^p \sum_{t=1}^{T-1} \sum_{k=1}^{K-1}
|\beta_{jtk}-\beta_{j(t+1)k}|. 
\end{equation*}
(The element notation in \eqref{eq::fusedmodel} emphasizes the fact
that the maximizing coefficients \smash{$(\hbeta_0,\hbeta)$} need not
be unique, since the log likelihood $\ell(\beta_0,\beta)$ need not be
strictly concave---e.g., this is the case when $p>n$.)

In broad terms, the lasso and fused lasso penalties encourage sparsity
and persistence, respectively, in the estimated coefficients
\smash{$\hbeta$}.  A larger value of the tuning parameter $\lambda_1
\geq 0$ generally corresponds to fewer nonzero entries in
\smash{$\hbeta$}; a larger value of the tuning parameter $\lambda_2
\geq 0$ generally corresponds to fewer change points in the piecewise
constant coefficient trajectories \smash{$\hbeta_{j \cdot k}$}, across
$t=1,\ldots T$.  We note that the form the log likelihood
$\ell(\beta_0,\beta)$ specified above assumes independence between the
outcomes across timepoints, which is a rather naive assumption given
the longitudinal nature of our problem setup. However, this naivety is
partly compensated by the role of the fused lasso penalty, which ties
together the multinomial models across timepoints. 

% In the following we will refer to the model presented in
% \eqref{eq::model} as the Logistic Fused Lasso model. 
% Depending on the value of $\lambda_1$ some elements of the minimizer  
% $\beta^*$ will be exactly 0. We will consider the variables
% associated to the non-zero elements of $\beta^*$ as the important
% predictors. The second term, the fused lasso penalty, represents a
% penalty on the difference of time consecutive coefficients
% $\beta_{jtk}$ and $\beta_{j(t+1)k}$. Depending on the value of
% $\lambda_2$, the minimizer $\beta^*$ will have piecewise constant
% elements in the index $t$ .
% The model is explained in detail in Section \ref{sec:model}. Several
% methods for the selection of $\lambda_1$ and $\lambda_2$ are
% presented in Section \ref{sec::selection}.\\ 

\begin{figure}[!ht]
\centering
\includegraphics[width=\textwidth]{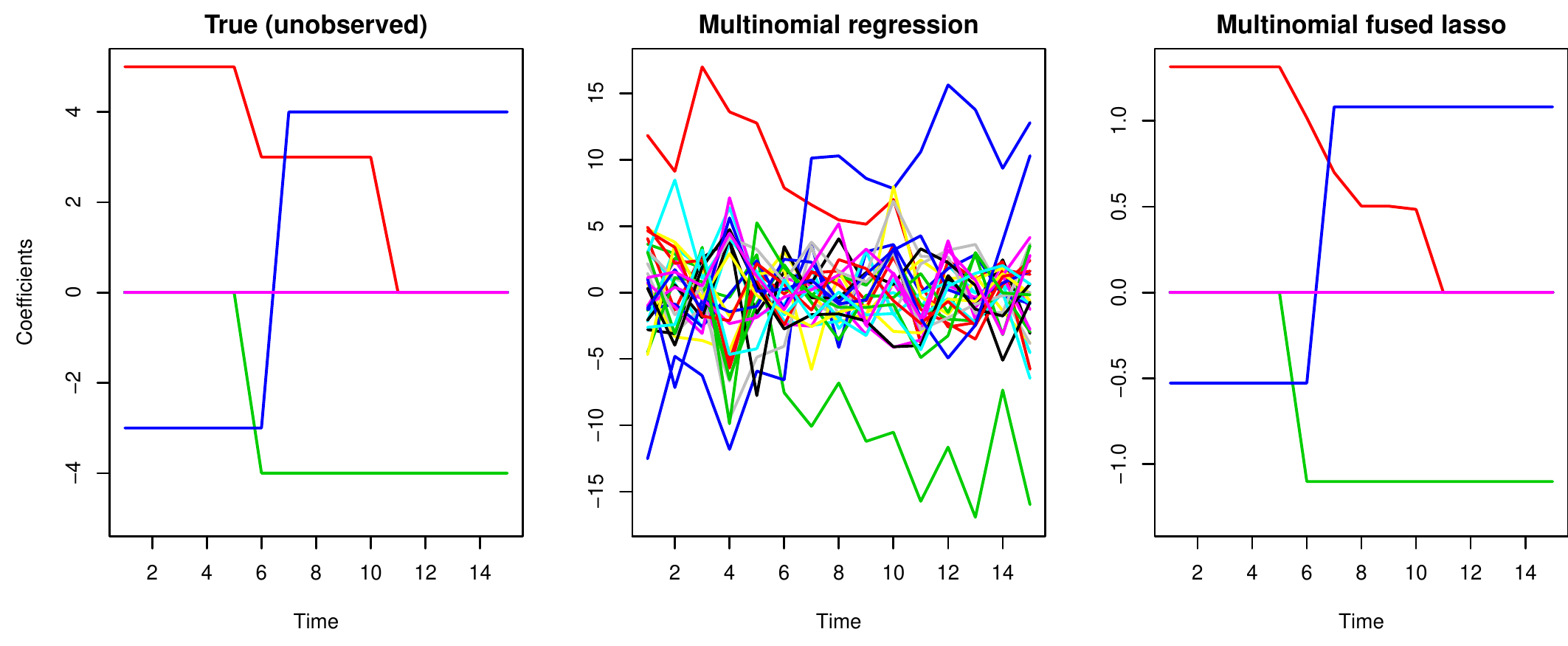}
\caption{A simple example with $n=50$, $T=15$, $K=2$, and $p=30$.  The
  left panel displays the true coefficent trajectories across
  timepoints $t=1,\ldots 15$ (only 3 of the 30 are nonzero); the
  middle panel shows the (unregularized) maximum likelihood estimates;
  the right panel shows the regularized estimates from
  \eqref{eq::multinomial}, with $\lambda_1=2.5$ and $\lambda_2=12.5$.} 
\label{fig::example}
\end{figure}

It helps to see an example. We consider a simple longitudinal problem
with $n=50$ individuals, $T=15$ timepoints, and $K=2$ classes.  At
each timepoint we sampled $p=30$ predictors independently from a
standard normal distribution. The true (unobserved) coefficient matrix
$\beta$ is now $30\times 15$; we set $\beta_{j\cdot}=0$ for $j=1\ldots
27$, and set the 3 remaining coefficients trajectories to be piecewise
constant across $t=1,\ldots 15$, as shown in the left panel of Figure
\ref{fig::example}.  In other words, the assumption here is that only 3 of
the 30 variables are relevant for predicting the outcome, and these
variables have piecewise constant effects over time.  We generated a
matrix of binary outcomes $Y$ according to the multinomial model
\eqref{eq::multinomial}, and computed the multinomial fused lasso
estimates \smash{$\hbeta_0,\hbeta$} in \eqref{eq::fusedmodel}.  The
right panel of Figure \ref{fig::example} displays these estimates (all
but the intercept \smash{$\hbeta_0$}) across $t=1,\ldots 15$, for a
favorable choice of tuning parameters $\lambda_1=2.5$,
$\lambda_2=12.5$; the middle plot shows the unregularized (maximum
likelihood) estimates corresponding to $\lambda_1=\lambda_2=0$.   

Each plot in Figure \ref{fig::example} has a $y$-axis that has been
scaled to suit its own dynamic range. 
We can see that the multinomial fused lasso estimates, with an
appropriate amount of regularization, pick up the underlying trend in
the true coefficients, though the overall magnitude of coefficients is
shrunken toward zero (an expected consequence of the $\ell_1$
penalties).  In comparison, the unregularized multinomial estimates
are wild and do not convey the proper structure.  From the perpsective
of prediction error, the multinomial fused lasso estimates offer a
clear advantage, as well: over 30 repetitions from the same simulation
setup, we used both the regularized coefficient estimates (with
$\lambda_1=2.5$ and $\lambda_2=12.5$) and the unregularized estimates
to predict the outcomes on an i.i.d.\ test set.  The average
prediction error using the regularized estimates was $0.114$ (with a
standard error of $0.014$), while the average prediction error from
the unregularized estimates was $0.243$ (with a standard error of
$0.022$).   

% On the right we see the result from the Logistic Fused Lasso model
% ($\lambda_1=2.5, \lambda_2= 12.5$). 27 estimated coefficients are
% correctly set to 0, while the remaining 3 reflect the same shape of
% the true unobserved coefficients, although they are shrunk towards
% 0. This is a consequence of the lasso and fused lasso penalties. 
% Finally we report the classification error on a test set. The entire
% procedure was repeated 30 times. 
% See Table \ref{tab::error}. 
% \begin{table}[h!]
% \begin{center}
% \begin{tabular}{| c | c | c |}
%  & Mean(Error) & sd(Error)  \\ \hline
% Logistic Regression & 0.240  & 0.023  \\ \hline
% Fused Lasso Model & 0.116 & 0.014 \\ \hline
% \end{tabular}
% \end{center}
% \caption{Misclassification Error for the prediction of the binary
% outcome $Y$ of Example \ref{example::toy}.} 
% \label{tab::error}
% \end{table}

\subsection{Related work and alternative approaches} 
\label{sec::related}

The fused lasso was first introduced in the statistics literature by
\citet{fuse}, and similar ideas based on total variation, starting
with \citet{tv}, have been proposed and studied extensively in the
signal processing community.  There have been many interesting
statistical applications of the fused lasso, in problems involving 
the analysis of comparative genomic hybridization data 
\citep{fusehot}, the modeling of genome association networks
\citep{graphguided}, and the prediction of colorectal cancer
\citep{lin2013}.  The fused lasso has in fact been applied to the
study of Alzheimer's disease in \citet{xin2014}, though these authors
consider a very different prediction problem than ours, based on
static magnetic resonance images, and do not have the time-varying
setup that we do.

Our primary motivation, which is the focus of Section 
\ref{sec::CHS}, is the problem of predicting the status of an
individual at age $t+10$ years from a number of variables measured 
at age $t$. 
For this we use the regularized multinomial model described in
\eqref{eq::multinomial}, \eqref{eq::fusedmodel}.  We encode
$K=3$ multinomial categories as normal, dementia, and
death: these are the three possible outcomes for any individual at
age $t+10$.  We are mainly interested in the prediction of
dementia; this task is complicated by the fact that risk factors 
for dementia are also known to be risk factors for death
\citep{rosvall2009apoe}, and so to account for this, we include
the death category in the multinomial classification model.  
An alternate approach would be
to use a Cox proportional hazards model \citep{cox1972}, where the
event of interest is the onset of dementia, and censorship corresponds
to  death. 

Traditionally, the Cox model is not fit with time-varying 
predictors or time-varying coefficients, but it can be naturally
extended to the setting considered in this work, even using the
same regularization schemes. Instead of the multinomial model
\eqref{eq::multinomial}, we would model the hazard function as 
\begin{equation}
\label{eq::hazard}
h(t+\Delta|X_{i \cdot t} = x) = h_0(t+\Delta) \cdot 
\exp(x^T \beta_{\cdot t}), 
\end{equation}
where $\beta \in \R^{p \times T}$ are a set of coefficients over time,
and $h_0$ is some baseline hazard function (that does not depend
on predictor measurements).  Note that the hazard model
\eqref{eq::hazard} relates the instantaneous rate of failure (onset of
dementia) at time $t+\Delta$ to the predictor measurements at time 
$t$.  This is as in the multinomial model \eqref{eq::multinomial},
which relates the outcomes at time $t+\Delta$ (dementia or death) to
predictor measurements at time $t$.  The coefficients in
\eqref{eq::hazard} would be determined by maximizing the 
partial log likelihood with the analogous lasso and fused lasso
penalties on $\beta$, as in the above multinomial setting
\eqref{eq::fusedmodel}.  

The partial likelihood approach can be viewed as a sequence of
conditional log odds models \citep{efron1977,kalb2002}, and therefore 
one might expect the (penalized) Cox regression model described here
to perform similarly to the (penalized) multinomial 
regression model pursued in this paper.  In fact, the computational
routine described in Section \ref{sec::GGD} would apply to the Cox
model with only very minor modifications (that concern the gradient
computations).  A rigorous comparison of the two approaches is beyond
the scope of the current manuscript, but is an interesting topic for
future development. 

\subsection{Outline}

The rest of this paper is organized as follows.  In Section
\ref{sec::GGD}, we describe a proximal gradient descent algorithm for
efficiently computing a solution \smash{$(\hbeta_0,\hbeta)$} in
\eqref{eq::multinomial}. 
% The basic algorithm is followed by several extensions that lead to
% the complete procedure of Algorithm % \ref{alg::FusedLassoComplete}.  
Next, we present an analysis of the CHS-CS data set in Section
\ref{sec::CHS}. Section \ref{sec::stability} discusses the stability
of estimated coefficients, and related concepts. In Section
\ref{sec::selection} we discuss numerous approaches for the selecting
the tuning parameters $\lambda_1,\lambda_2 \geq 0$ that govern the
strength of the lasso and fused lasso penalties in
\eqref{eq::multinomial}.    In Section \ref{sec::discussion}, we
conclude with some final comments and lay out ideas for future work.

\section{A proximal gradient descent approach}
\label{sec::GGD}
In this section, we describe an efficient proximal gradient descent
algorithm for computing solutions of the fused lasso regularized
multinomial regression problem \eqref{eq::fusedmodel}. While a number
of other algorithmic approaches are possible, such as implementations
of the alternating direction method of multipliers \citep{admm}, we
settle on the proximal gradient method because of its simplicity, and 
because of the extremely efficient, direct proximal mapping
associated with the fused lasso regularizer.
We begin by reviewing proximal gradient descent in generality, then we
describe its implementation for our problem, and a number of practical 
considerations like the choice of step size, and stopping criterion.  

\subsection{Proximal gradient descent}
\label{sec::proxintro}

Suppose that $g : \R^d \rightarrow \R$ is convex and differentiable, 
$h : \R^d \rightarrow \R$ is convex, and we are interested in
computing a solution 
\begin{equation*}
x^\star \in \argmin_{x \in \R^d} \; g(x) + h(x).
\end{equation*}
If $h$ were assumed differentiable, then the criterion
$f(x)=g(x)+h(x)$ is convex and differentiable, and repeating the
simple gradient descent steps 
\begin{equation}
\label{eq::gradstep}
x^+ = x - \tau \nabla f(x)
\end{equation}
suffices to minimize $f$, for an appropriate choice of step size
$\tau$.  (In the above, we write $x^+$ to denote the gradient descent
update from the current iterate $x$.)  If $h$ is not  
differentiable, then gradient descent obviously does not apply, but
as long as $h$ is ``simple'' (to be made precise
shortly), we can apply a variant of gradient descent that shares many
of its properties, called {\it proximal gradient descent}. 
Proximal gradient descent is often also called composite or
generalized gradient descent, and in this routine we repeat the steps 
\begin{equation}
\label{eq::proxstep}
x^+ = \prox_{h,\tau} \big( x - \tau \nabla g(x) \big)
\end{equation}
until convergence, where $\prox_{h,\tau} : \R^d \rightarrow \R^d$ is
the proximal mapping associated with $h$ (and $\tau$),  
\begin{equation}
\label{eq::proxmap}
\prox_{h,\tau}(x) = \argmin_{z \in \R^d} \;
\frac{1}{2\tau} \|x-z\|_2^2 + h(z).
\end{equation}
(Strict convexity of the above criterion ensures that it has a unique
minimizer, so that the proximal mapping is well-defined.)  Provided
that $h$ is simple, by which we mean that its proximal map
\eqref{eq::proxmap} is explicitly computable, the proximal gradient
descent steps \eqref{eq::proxstep} are straightforward and resemble the
classical gradient descent analogues \eqref{eq::gradstep}; we
simply take a gradient step in the direction governed by the smooth
part $g$, and then apply the proximal map of $h$.  A slightly more
formal perspective argues that the updates \eqref{eq::proxmap} are the 
result of minimizing $h$ plus a quadratic expansion of $g$, around the
current iterate $x$.

Proximal gradient descent has become a very popular tool for
optimization problems in statistics and machine learning, where
typically $g$ represents a smooth loss function, and $h$ a nonsmooth 
regularizer.  This trend is somewhat recent, even
though the study of proximal mappings has a long history of
in the optimization community (e.g., see \citet{prox} for a nice
review paper).  In terms of convergence properties, proximal gradient
descent enjoys essentially the same convergence rates as gradient
descent under the analogous assumptions, and is amenable to
acceleration techniques just like gradient descent 
(e.g., \citet{nestcomp}, \citet{fista}).  
Of course, for proximal gradient
descent to be applicable in practice, one must be able to exactly (or
even approximately) compute the proximal map of $h$ in
\eqref{eq::proxmap}; fortunately, this is possible for many
optimization problems, i.e., many common regularizers $h$,
that are encountered in statistics. In our case, the proximal mapping
reduces to solving a problem of the form 
\begin{equation}
\label{eq::1dfused}
\hat{\theta} = \argmin_{\theta} \frac{1}{2} \|x-\theta\|_2^2 + 
\lambda_1 \sum_{i=1}^m |\theta_i| +
\lambda_2 \sum_{i=1}^{m-1} |\theta_i-\theta_{i+1}|.
\end{equation}
This is often called the fused lasso signal approximator (FLSA)
problem,
%1-dimensional (Gaussian) fused lasso problem, 
and extremely fast, linear-time algorithms exist to compute its
solution.  In particular, we rely on an elegant dynamic
programming approach proposed by \citet{nickdp}.

% \begin{algorithm}
% \begin{algorithmic}[1]
% 	  \INPUT $\beta^{(0)}, \tau, S$
% 	  \OUTPUT $\beta^*$
% 	    \For{$s=1 \text{ to } S$} 
% 		\State set $\beta^{(s)}= \text{prox}_{\tau} \left( [\beta^{(s-1)}- \tau \nabla g(\beta^{(s-1)})] \right)$
% 	\EndFor
% \State $\beta^* \leftarrow \beta^{(S)}$
% \State \textbf{return} $\beta^*$
% \end{algorithmic}
% \caption{Proximal Gradient Descent}
% \label{alg::ggd}
% \end{algorithm}

\subsection{Application to the multinomial fused lasso problem}
\label{sec::proxapp}

The problem in \eqref{eq::fusedmodel} fits into the desired form for
proximal gradient descent, with $g$ the multinomial regression loss
(i.e., negative multinomial regression log likelihood) and $h$ the
lasso plus fused lasso penalties.  Formally, we can rewrite
\eqref{eq::fusedmodel} as 
\begin{equation}
\label{eq::fusedmodel2}
(\hbeta_0, \hbeta ) \in 
\argmin_{\beta_0,\beta} \; g(\beta_0, \beta) + h(\beta_0,\beta), 
\end{equation}
where $g$ is the convex, smooth function
\begin{equation*}
g(\beta_0, \beta) = \sum_{t=1}^T \sum_{i=1}^{n} 
\left\{
  \sum_{k=1}^{K-1} - \mathbb{I}(Y_{it}=k) 
  (\beta_{0tk}+X_{i\cdot t}\beta_{\cdot tk}) + 
  \log \left(1+\sum_{h=1}^{K-1} \exp(\beta_{0th} + X_{i\cdot
    t}\beta_{\cdot th}) \right) \right\},
\end{equation*}
%(we write $\mathbb{I}(\cdot)$ for the indicator function), 
and $h$ is the convex, nonsmooth function
\begin{equation*}
h(\beta_0,\beta) = \lambda_1 \sum_{j=1}^p \sum_{t=1}^T
\sum_{k=1}^{K-1}  |\beta_{jtk}| + \lambda_2 \sum_{j=1}^p
\sum_{t=1}^{T-1} \sum_{k=1}^{K-1} |\beta_{jtk}-\beta_{j(t+1)k}| .  
\end{equation*}
Here we consider fixed values $\lambda_1,\lambda_2 \geq 0$. As
described previously, each of these tuning parameters will have a big
influence on the strength of their respective penalty terms, and hence
the properties of the computed estimate \smash{$(\hbeta_0,\hbeta)$};
we discuss the selection of $\lambda_1$ and $\lambda_2$ in
Section \ref{sec::selection}.  We note that the intercept coefficients
$\beta_0$ are not penalized.

To compute the proximal gradient updates, as given in
\eqref{eq::proxstep}, we must consider two quantities: the
gradient of $g$, and the proximal map of $h$.  First, we discuss the 
gradient. As $\beta_0 \in \R^{T\times (K-1)}$, 
$\beta \in \R^{p\times T \times (K-1)}$, we may consider the gradient
as having dimension 
$\nabla g(\beta_0,\beta) \in \R^{(p+1) \times T \times (K-1)}$.  We
will index this as $[\nabla g(\beta_0,\beta)]_{jtk}$ for $j=0,\ldots  
p$, $t=1,\ldots T$, $k=1,\ldots K-1$; hence note that $[\nabla 
g(\beta_0,\beta)]_{0tk}$ gives the partial derivative of $g$ with
respect to $\beta_{0tk}$, and $[\nabla g(\beta_0,\beta)]_{jtk}$ the
partial derivative with respect to $\beta_{jtk}$, for $j=1,\ldots p$.   
% The minimization problem, in the form given in \eqref{eq:GeneralProblem} is
% \begin{align*}
% (\beta_0^+, \beta^+) = \argmin_{\substack{ B_0 \in M_{T,(K-1)} \\ B \in A_{P,T,(K-1)}}} & \sum_{t=1}^T \sum_{k=1}^{K-1}  \frac{1}{2} \left[ B_{0tk} - \left(\beta_{0tk} - \tau [\nabla g(\beta_0,\beta)]_{0tk} \right) \right]^2 +\\ 
% + \sum_{j=1}^P\sum_{t=1}^T  \sum_{k=1}^{K-1} &\left\{ \frac{1}{2} \left[ B_{jtk} - \left(\beta_{jtk} - \tau [\nabla g(\beta_0,\beta)]_{jtk} \right) \right]^2 + \tau \lambda_1  |B_{jtk}| + \tau \lambda_2 |B_{jtk}-B_{j(t+1)k}| \right\} ,
% \end{align*}
% where the gradient $\nabla g(\beta_0,\beta) \in A_{(P+1),T,(K-1)}$ is
% defined element by element. 
For generic $t,k$, we have
\begin{equation}
\label{eq::gradient0}
[\nabla g(\beta_0,\beta)]_{0tk}
%=\left.\frac{\partial
%    g(\beta_0,\beta)}{\partial \beta_{0tk}}\right|_{(\beta_0,\beta)}
=
\sum_{i=1}^{n} \left( -\mathbb{I}(Y_{it}=k) +
\frac{ \exp(\beta_{0tk} + X_{i\cdot t}
\beta_{\cdot tk})}
{1+\sum_{h=1}^{K-1} \exp(\beta_{0th} + 
X_{i\cdot t}\beta_{\cdot th})}\right),
\end{equation}
and for $j \geq 1$,
\label{eq::gradient}
\begin{equation}
[\nabla g(\beta_0,\beta)]_{jtk}
%=\left.\frac{\partial
%    g(\beta_0,\beta)}{\partial \beta_{jtk}}\right|_{(\beta_0,\beta)}
=  \sum_{i=1}^{n} 
\left( -\mathbb{I}(Y_{it}=k)  X_{ijt} +  
X_{ijt}\frac{ \exp(\beta_{0tk} +
X_{i\cdot t}\beta_{\cdot tk})}
{1+\sum_{h=1}^{K-1} \exp(\beta_{0th} +
X_{i\cdot t}\beta_{\cdot th})}\right). 
\end{equation}
It is evident that computation of the gradient requires $O(npTK)$.

Now, we discuss the proximal operator.  Since the intercept
coefficients $\beta_0 \in \R^{T\times (K-1)}$ are left unpenalized,
the proximal map over $\beta_0$ just reduces to the identity, and the
intercept terms undergo the updates
\begin{equation*}
\beta_{0tk}^+ = \beta_{0tk} - \tau [\nabla g(\beta_0,\beta)]_{0tk} 
\;\;\;\text{for}\;\, t=1,\ldots T, \; k=1, \ldots K-1. 
\end{equation*}
Hence we consider the proximal map over $\beta$ alone.   
At an arbitrary input $x \in \R^{p \times T \times (K-1)}$, this is 
\begin{equation*}
\argmin_{z \in \R^{p\times T \times (K-1)}} \;
\frac{1}{2\tau} \sum_{j=1}^p \sum_{t=1}^T
\sum_{k=1}^{K-1} (x_{jtk} - z_{jtk})^2 +
 \lambda_1 \sum_{j=1}^p \sum_{t=1}^T
\sum_{k=1}^{K-1}  |\beta_{jtk}| + \lambda_2 \sum_{j=1}^p
\sum_{t=1}^{T-1} \sum_{k=1}^{K-1} |\beta_{jtk}-\beta_{j(t+1)k}|, 
\end{equation*}
which we can see decouples into $p(K-1)$ separate minimizations, one
for each predictor $j=1,\ldots p$ and class $k=1,\ldots K-1$.  In
other words, the coefficients $\beta$ undergo the updates
\begin{multline}
\label{eq::bupdate}
\beta^+_{j\cdot k} =
 \argmin_{\theta \in \R^T} \frac{1}{2}
 \sum_{t=1}^T \Big(\big(\beta_{j\cdot k} - \tau 
[\nabla g(\beta_0,\beta)]_{j\cdot k}\big) - \theta \Big)^2 + 
  \tau \lambda_1  \sum_{t=1}^T |\theta_t| +\tau \lambda_2
  \sum_{t=1}^{T-1} |\theta_t-\theta_{t+1}|,\\ 
\text{for}\;\, j=1,\ldots p, \; k=1,\ldots K-1,
\end{multline}
each minimization being a fused lasso signal approximator problem
%1-dimensional Gaussian fused lasso problem
\citep{fuse}, i.e., of the form \eqref{eq::1dfused}.  There are many
computational approaches that may be applied to such a problem
structure; we employ a specialized, highly efficient algorithm by
\citet{nickdp} that is based on dynamic programming.  This algorithm
requires $O(T)$ operations for each of the problems in
\eqref{eq::bupdate}, making the total cost of the update $O(pTK)$
operations.  Note that this is actually dwarfed by the cost of
computing the gradient $\nabla g(\beta_0,\beta)$ in the first place,
and therefore the total complexity of a single iteration of our proposed
proximal gradient descent algorithm is $O(npTK)$.  

% \begin{algorithm}
% \begin{algorithmic}[1]
% 	  \INPUT $X,Y,\beta_0^{(0)},\beta^{(0)}, \lambda_1, \lambda_2, \tau, S$
% 	  \OUTPUT $(\hat\beta_0^*,\hat\beta^*)$
% 	    \For{$s=1 \text{ to } S$} \\
% 	Update the intercept: $\beta_{0\cdot \cdot}^{(s)} \leftarrow \beta^{(s-1)}_{0\cdot \cdot} - \tau [\nabla g(\beta_0^{(s-1)},\beta^{(s-1)})]_{0\cdot\cdot}$
% 	    \For{$j=1 \text{ to } P$}
% 			\For{$k=1 \text{ to } (K-1)$} 
% 			 \State set $\beta^{(s)}_{j\cdot k} \leftarrow \text{prox}_\tau (\beta^{(s-1)}_{j \cdot k} -\tau [\nabla g(\beta_0^{(s-1)},\beta^{(s-1)})]_{j\cdot k})$ 
% 			\EndFor
% 		\EndFor
% 	\EndFor
% \State $\hat\beta_0^* \leftarrow \beta_0^{(S)}$
% \State $\hat\beta^* \leftarrow \beta^{(S)}$
% \State \textbf{return} $(\hat\beta_0^*, \hat\beta^*)$
% \end{algorithmic}
% \caption{Basic Proximal Gradient Descent for the Fused Lasso Model}
% \label{alg::FusedLasso}
% \end{algorithm}

\subsection{Practical considerations}
\label{sec::practical}

We discuss several practical issues that arise in applying the
proximal gradient descent algorithm.
% Several extensions of the Proximal Gradient Descent algorithm of
% Algorithm \ref{alg::FusedLasso} are possible. In this section we will
% see that the step size $\tau$ of steps 2 and 5 can be selected at each
% iteration, using a strategy known as backtracking line search. We will
% also present two criterions to terminate the algorithm when the
% current value of the coefficient is sufficiently close to the
% minimizer $\beta^*$. Before presenting the two extensions we propose a
% solution for the problem of missing values in the outcome $Y$, which
% could lead to an unbalanced penalty of some of the coefficients that
% we are trying to estimate.

\subsubsection{Backtracking line search}

Returning to the generic perpsective for proximal gradient descent as
described in Section \ref{sec::proxintro}, we rewrite the proximal
gradient descent update in \eqref{eq::proxstep} as
\begin{equation}
\label{eq::genstep}
x^+ = x - \tau G_\tau (x),
\end{equation}
where $G_\tau(x)$ is called the {\it generalized gradient} and is
defined as
\begin{equation*}
G_\tau = \frac{x - \prox_{h,\tau}(x-\tau\nabla g(x))}{\tau}.
\end{equation*}
The update is rewritten in this way so that it more closely resembles
the usual gradient update in \eqref{eq::gradstep}.  We can see that, 
analogous to the gradient descent case, the choice of parameter
$\tau>0$ in 
each iteration of proximal gradient descent determines the magnitude
of the update in the direction of the generalized gradient
$G_\tau(x)$.  Classical analysis shows that if $\nabla g$ is
Lipschitz with constant $L>0$, then proximal gradient descent
converges with any fixed choice of step size $\tau \leq 1/L$
across all iterations.  In most practical situations, however, the
Lipschitz constant $L$ of $\nabla g$ is not known or easily
computable, and we rely on an adaptive scheme for choosing an
appropriate step
size at each iteration; backtracking line search is one such scheme,
which is straightforward to implement in practice and guarantees
convergence of the algorithm under the same Lipschitz assumption on
$\nabla g$ (but importantly, without having to know its Lipschitz
constant $L$).  Given a shrinkage factor $0 < \gamma < 1$, the
backtracking line search routine at a
given iteration of proximal gradient descent starts with $\tau=\tau_0$
(a large initial guess for the step size), and while
\begin{equation}
\label{eq::backtrack}
g\big(x-\tau G_\tau(x)\big) > g(x) - \tau \nabla g(x)^T G_\tau(x) + 
\frac{\tau}{2}\|G_\tau(x)\|_2^2, 
\end{equation}
it shrinks the step size by letting $\tau=\gamma \tau$.  Once the exit 
criterion is achieved (i.e., the above is no longer satisfied), the
proximal gradient descent algorithm then uses the current value of
$\tau$ to take an update step, as in \eqref{eq::genstep} (or
\eqref{eq::proxstep}). 

In the case of the multinomial fused lasso problem, the generalized
gradient is of dimension $G_\tau(\beta_0,\beta) \in 
\R^{(p+1)\times T\times (K-1)}$, where  
\begin{equation*}
[G_\tau(\beta_0,\beta)]_{0\cdot\cdot} = 
[\nabla g(\beta_0,\beta)]_{0\cdot\cdot},
\end{equation*}
and
\begin{equation*}
[G_\tau(\beta_0,\beta)]_{j\cdot k} = 
\frac{\beta_{j\cdot k} - \prox_{\mathrm{FLSA},\tau}
(\beta_{j \cdot k} -\tau [\nabla g(\beta_0,\beta)]_{j\cdot k})}
{\tau} \;\;\; \text{for} \;\, 
j=1,\ldots p, \;k=1,\ldots K-1.
\end{equation*}
Here $\prox_{\mathrm{FLSA},\tau}
(\beta_{j \cdot k} -\tau [\nabla g(\beta_0,\beta)]_{j\cdot k})$
is the proximal map defined by the fused lasso signal approximator
evaluated at $\beta_{j \cdot k} -\tau [\nabla
g(\beta_0,\beta)]_{j\cdot k}$, i.e., the right-hand side in
\eqref{eq::bupdate}.  Backtracking line search now applies just as
described above. %See Algorithm \ref{alg::backtracking}.

\subsubsection{Stopping criteria}

The simplest implementation of proximal gradient descent would 
run the algorithm for a fixed, large number of steps $S$.  A more
refined approach would check a stopping criterion at the end of each
step, and terminate if such a criterion is met.  Given a
tolerance level $\epsilon>0$, two common stopping criteria are then
based on the relative difference in function values, as in 
\begin{equation*}
\text{stopping criterion 1: terminate if}\;\
 C_1 = \frac{| f(\beta_0^+,\beta^+) - f(\beta_0,\beta) |}
{f(\beta_0,\beta)} \leq \epsilon,
\end{equation*}
and the relative difference in iterates, as in
\begin{equation*}
\text{stopping criterion 2: terminate if}\;\,
C_2 = \frac{\|( \beta_0^+,\beta^+) - (\beta_0,\beta)\|_2}
{\|(\beta_0,\beta)\|_2} \leq \epsilon.
\end{equation*}
The second stopping criterion is generally more stringent, and may be
hard to meet in large problems, given a small tolerance $\epsilon$.

For the sake of completeness, we outline the full proximal gradient
descent procedure in the notation of the multinomial fused lasso
problem, with backtracking line search and the first stopping
criterion, in Algorithms \ref{alg::proxgrad} and \ref{alg::backtracking} below. 

\begin{algorithm}[!htb]
\begin{algorithmic}[1]
\INPUT Predictors $X$, outcomes $Y$, tuning parameter values
$\lambda_1,\lambda_2$, initial coefficient guesses 
\smash{$(\beta_0^{(0)},\beta^{(0)})$}, maximum number of iterations $S$, 
initial step size before backtracking $\tau_0$, backtracking shrinkage
parameter $\gamma$, tolerance $\epsilon$
\OUTPUT Approximate solution \smash{$(\hbeta_0,\hbeta)$} 
\State $s = 1$, $C=\infty$
\While{($s \leq S$ and $C > \epsilon$)} 
\State Find $\tau_s$ using backtracking, 
Algorithm \ref{alg::backtracking} 
(INPUT: \smash{$\beta_0^{(s-1)}, \beta^{(s-1)}, \tau_0, \gamma$})  
\State Update the intercept: 
\smash{$\beta_{0\cdot \cdot}^{(s)} =
\beta^{(s-1)}_{0\cdot \cdot} - 
\tau_s [\nabla g(\beta_0^{(s-1)},\beta^{(s-1)})]_{0\cdot\cdot}$}
\For{$j=1,\ldots p$}
\For{$k=1,\ldots (K-1)$} 
\State Update 
\smash{$\beta^{(s)}_{j\cdot k} = \prox_{\mathrm{FLSA},\tau_s}
(\beta^{(s-1)}_{j \cdot k} -\tau_s [\nabla
g(\beta_0^{(s-1)},\beta^{(s-1)})]_{j\cdot k})$}
\EndFor
\EndFor
\State Increment $s = s+1$
\State Compute \smash{$C = 
[f(\beta_0^{(s)},\beta^{(s)}) - f(\beta_0^{(s-1)},\beta^{(s-1)})]/
f(\beta_0^{(s-1)},\beta^{(s-1)})$}
\EndWhile
\State 
\smash{$\hbeta_0 = \beta_0^{(s)}$, $\hbeta = \beta^{(s)}$} \\ 
\Return \smash{$(\beta_0, \hbeta)$}
\end{algorithmic}
\caption{Proximal gradient descent for the multinomial fused lasso}
\label{alg::proxgrad}
\end{algorithm}

\begin{algorithm}[!htb]
\begin{algorithmic}[1]
  \INPUT $\beta_0,\beta, \tau_0,\gamma$
  \OUTPUT $\tau$
  \State $\tau = \tau_0$
  \While{(true)}
\State Compute
$[G_\tau(\beta_0,\beta)]_{0\cdot\cdot} = 
[\nabla g(\beta_0,\beta)]_{0\cdot\cdot}$
\For{$j=1,\ldots p$}
\For{$k=1,\ldots (K-1)$} 
\State Compute 
$[G_\tau(\beta_0,\beta)]_{j\cdot k} = 
[\beta_{j\cdot k} - \prox_{\mathrm{FLSA},\tau}
(\beta_{j \cdot k} -\tau [\nabla g(\beta_0,\beta)]_{j\cdot k})]/
\tau$
\EndFor
\EndFor
\If{$g((\beta_0,\beta) -\tau G_\tau(\beta_0,\beta)) > 
g(\beta_0,\beta) - \tau [\nabla g(\beta_0,\beta)]^T
G_\tau(\beta_0,\beta) + \frac{\tau}{2}
\|G_\tau(\beta_0,\beta)\|_2^2$}
\State Break
\Else
\State Shrink $\tau = \gamma \tau$
\EndIf
\EndWhile \\
\Return $\tau$
\end{algorithmic}
\caption{Backtracking line search for the multinomial fused lasso}
\label{alg::backtracking}
\end{algorithm}

\subsubsection{Missing individuals}
\label{sec::missingIndiv}
Often in practice, some individuals are not present at some
timepoints in the longitudinal study, meaning that one or both of
their outcome values and predictor measurements are missing over 
a subset of $t=1,\ldots T$.  Let $I_t$ denote the set of
completely observed individuals (i.e., with both predictor
measurements and outcomes observed) at time $t$, and let $n_t=|I_t|$.
The simplest strategy to accomodate such missingness would be to
compute the loss function $g$ only observed individuals, so that   
\begin{equation*}
g(\beta_0, \beta) = \sum_{t=1}^T \sum_{i \in I_t} 
\left\{
  \sum_{k=1}^{K-1} - \mathbb{I}(Y_{it}=k) 
  (\beta_{0tk}+X_{i\cdot t}\beta_{\cdot tk}) + 
  \log \left(1+\sum_{h=1}^{K-1} \exp(\beta_{0th} + X_{i\cdot
    t}\beta_{\cdot th}) \right) \right\}.
\end{equation*}
An issue arises when the effective sample size $n_t$ is quite variable
across timepoints $t$: in this case, the penalty terms can have quite
different effects on the 
coefficients $\beta_{\cdot\cdot t}$ at one time $t$ versus
another. That is, the coefficients $\beta_{\cdot\cdot t}$ at a time
$t$ in which $n_t$ is small experience a relatively small loss term    
\begin{equation}
\label{eq::gt}
\sum_{i \in I_t} 
\left\{
  \sum_{k=1}^{K-1} - \mathbb{I}(Y_{it}=k) 
  (\beta_{0tk}+X_{i\cdot t}\beta_{\cdot tk}) + 
  \log \left(1+\sum_{h=1}^{K-1} \exp(\beta_{0th} + X_{i\cdot
    t}\beta_{\cdot th}) \right) \right\},
\end{equation}
simply because there are fewer terms in the above sum compared to a
time with a larger effective sample size; however, the penalty term 
\begin{equation*}
\lambda_1 \sum_{j=1}^p 
\sum_{k=1}^{K-1}  |\beta_{jtk}| + \lambda_2 \sum_{j=1}^p
\sum_{k=1}^{K-1} |\beta_{jtk}-\beta_{j(t+1)k}| 
\end{equation*}
remains comparable across all timepoints, regardless of sample size.
A fix would be to scale the loss term in \eqref{eq::gt} by $n_t$ to
make it (roughly) independent of the effective sample size, so that
the total loss becomes
\begin{equation}
\label{eq:gnew}
g(\beta_0,\beta) = 
\sum_{t=1}^T
\frac{1}{n_t}
\sum_{i \in I_t} 
\left\{
  \sum_{k=1}^{K-1} - \mathbb{I}(Y_{it}=k) 
  (\beta_{0tk}+X_{i\cdot t}\beta_{\cdot tk}) + 
  \log \left(1+\sum_{h=1}^{K-1} \exp(\beta_{0th} + X_{i\cdot
    t}\beta_{\cdot th}) \right) \right\}.
\end{equation}
This modification indeed ends up being important for the
Alzheimer's analysis that we present in Section \ref{sec::CHS}, since
this study has a number of individuals in the tens at some timepoints,
and in the hundreds for others.  The proximal gradient descent
algorithm described in this section extends to cover the loss in
\eqref{eq:gnew} with only trivial modifications.

% \begin{equation}
% [\nabla g(\beta_0,\beta)]_{jtk}=\frac{1}{|N_t|}\sum_{i \in N_t} \left[  - \mathbb{I}(Y_{it}=k)  x_{ijt}  + x_{ijt}\frac{ \exp(\beta_{0tk} +X_{i\cdot t}\beta_{\cdot tk})}{\left[1+\sum_{h=1}^{K-1} \exp(\beta_{0th} +X_{i\cdot t}\beta_{\cdot th})\right]}\right]
% \label{eq::gradient0Bis}
% \end{equation}
% and
% \begin{equation}
% [\nabla g(\beta_0,\beta)]_{0tk}= \frac{1}{|N_t|}\sum_{i \in N_t} \left[  - \mathbb{I}(Y_{it}=k)   +\frac{ \exp(\beta_{0tk} + X_{i\cdot t}\beta_{\cdot tk})}{\left[1+\sum_{h=1}^{K-1} \exp(\beta_{0th} + X_{i\cdot t}\beta_{\cdot th})\right]}\right].
% \label{eq::gradientBis}
% \end{equation}

\subsection{Implementation in C++ and R}

An efficient C++ implementation of the proximal gradient descent
algorithm described in this section, with an easy interface to R, is
available from the second author's website:
\url{http://www.stat.cmu.edu/~flecci}.  In the future, this will be
available as part of the R package {\tt glmgen}, which broadly fits
generalized linear models under generalized lasso regularization.

\section{Alzheimer's Disease data analysis}
\label{sec::CHS}

In this section, we apply the proposed estimation method to the data
of the the Cardiovascular Health Study Cognition Study (CHS-CS), a
rich database of thousands of multiple cognitive, metabolic, 
cardiovascular, cerebrovascular, and neuroimaging variables obtained
over the past 24 years for people of ages 65 to 110 years old
\citep{fried1991cardiovascular, lopez2007incidence}.  

The complex relationships between age and other risk factors produce
highly variable natural histories from normal cognition to the
clinical expression of Alzheimer's disease, either as dementia or its
prodromal syndrome, mild cognitive impairment (MCI)
\citep{lopez2003prevalence, saxton2004preclinical, lopez2007incidence,
  sweet2012effect, 2014mixed}. 
Many studies involving the CHS-CS data have shown the importance of a
range of risk factors in predicting the time of onset of clinical
dementia. The risk of dementia is affected by the presence of the
APOE*4 allele, male sex, lower education, and having a family history
of dementia \citep{fitzpatrick2004incidence, tang1996relative,
  launer1999rates}. Medical risks include the presence of systemic
hypertension, diabetes mellitus, and cardiovascular or cerebrovascular
disease \citep{kuller2003risk, irie2005type, skoog199615}. Lifestyle
factors affecting risk include physical and cognitive activity, and
diet \citep{verghese2003leisure, erickson2010physical,
  scarmeas2006mediterranean}. 

A wide range of statistical approaches has been considered in these
studies, including exploratory statistical summaries, hypothesis
tests, survival analyses, logistic regression models, and latent
trajectory models. 
None of these methods can directly accommodate a large number of
predictors that can potentially exceed the number of observations. 
A small number of variables was often chosen a priori to match the
requirements of a particular model, neglecting the full potential  
of the CHS-CS data, which consists of thousands of variables. 

The approach that we introduced in Section \ref{sec::introduction} can
accommodate an array of predictors of arbitrary dimension, using
regularization to maintain a well-defined predictive model and avoid
overfitting.  Our goal is to identify important risk factors for
the prediction of the cognitive status at $t+10$ years of age 
($\Delta=10$), given predictor measurements at $t$ years of age, for
$t=65,66, \ldots ,98$. 
We use the penalized log likelihood criterion in
\eqref{eq::fusedmodel} to estimate the coefficients of the multinomial
logit model in \eqref{eq::multinomial}. The lasso penalty forces the
solution to be sparse, allowing us to identify a few important
predictors among the thousands of variables of the CHS-CS data. 
% The second property of our solution is the persistence of active
% predictors: 
The fused lasso penalty allows for a few change points in
the piecewise constant coefficient trajectories 
\smash{$\hbeta_{j \cdot k}$}, across $t=65,\ldots 98$. Justification 
for this second penalty is based on the scientific intuition that
predictors that are clinically important should have similar effects
in successive ages.  

% This set-up allows us to understand the interaction of risk factors
% with age, which is based on a major assumption that the age itself is
% less important as a risk modifier than the various medical factors
% that change as a consequence of aging
% \citep{rabbit1991courseAndcauses}. \todo{is this true?} 

\subsection{Data preprocessing}

We use data from the $n=924$ individuals in the Pittsburgh 
section of the CHS-CS, recorded between 1990 and 2012. 
Each individual underwent clinical and cognitive assessments at
multiple ages, all falling in the range $65, \ldots 108$. 
The matrix of (future) outcomes $Y$ has dimension $n \times 34$:  
for $i=1,\ldots 924$ and $t=65,\ldots 98$, the outcome $Y_{it}$
stores the cognitive status at age $t+10$ and can assume one of the  
following values: 
\begin{equation*}
Y_{it}=
\begin{cases}
1 & \text{if normal}\\
2 & \text{if MCI/dementia}\\
3 & \text{if dead}
\end{cases}.
\end{equation*}
MCI is included in the same class as dementia, 
as they are both instances of cognitive impairment.
Hence the proposed multinomial model predicts the onset of
MCI/dementia, in the presence of a separate death category.  
This is done to implicitly adjust for the confounding effect of death,
as some risk factors for dementia are also known to be risk factors
for death \citep{rosvall2009apoe}.   

The array of predictors $X$ is composed of time-varying variables that
were recorded at least twice during the CHS-CS study, and
time-invariant variables, such as gender and race. A complication in 
the data set is the ample amount of missingness in the array of
predictors. We impute missing values using a uniform rule for all
possible causes of missingness. A missing value at age $t$ is imputed
by taking the closest past measurement from the same individual, if
present. If all the past values are missing, the global median from
people of age $t$ is used. The only exception is the case of
time-invariant predictors, whose missing values are imputed by either
future or past values, as available.  

Categorical variables with $m$ possible outcomes are converted to
$m-1$ binary variables and all the predictors are standardized to have
zero mean and unit standard deviation. This is a standard procedure in
regularization, as the lasso and fused lasso penalties puts
constraints on the size of the coefficients associated with each
variable \citep{tibshirani1997lasso}.  
To be precise, imputation of missing values and standardization of the
predictors are performed within each of the folds used in the
cross-validation method for the choice of the tuning parameters
$\lambda_1$ and $\lambda_2$ (discussed below), and then again for the
full data set in the final estimation procedure that uses the selected 
tuning parameters. 

The final array of predictors $X$ has dimension $924 \times 1050
\times 34$, where $1050$ is the number of variables recorded over the
period of 34 years of age range.  

\subsection{Model and algorithm specification}

In the Alzheimer's Disease application, the multinomial model 
in \eqref{eq::multinomial} is determined by two equations, as there
are three possible outcomes (normal, MCI/ dementia, death); the
outcome ``normal'' is taken as the base class. We will refer to the
two equations (and the corresponding sets of coefficients) as the 
``dementia vs normal'' and ``death vs normal'' equations,
respectively.   

We use the proximal gradient descent algorithm described in Section
\ref{sec::GGD} to estimate the coefficients  
that maximize the penalized log likelihood criterion in
\eqref{eq::fusedmodel}. The initializations \smash{$(\beta_0^{(0)},
  \beta^{(0)})$} are set to be zero matrices, the maximum number of
iterations is $S=80$, the initial step size before backtracking is
$\tau_0=20$, the backtracking shrinkage parameter is $\gamma=0.6$ and
the tolerance of the first stopping criterion (relative difference in
function values) is $\epsilon=0.001$. 
We select the tuning parameters by a 4-fold cross-validation procedure
that minimizes the misclassification error. 
The selected parameters are $\lambda_1 = 0.019$ and $\lambda_2 =
0.072$, which yield an average prediction error of 0.316 (standard
error 0.009).  
%\todo{with 580 degrees of freedom} 
Section \ref{sec::selection} discusses more details on the model
selection problem.

The number $n_t$ of outcomes observed at age $t$ varies across time, 
for two reasons: first, different subjects entered the study at
different ages, and second, once a subject dies at time $t_0$, we
exclude them consideration in the model formed at all ages $t>t_0$, to 
predict the outcomes of individuals at age $t+10$.
% As people entered the study at different ages, the number $n_t$ of
% outcomes observed at ages $t$ varies in the matrix $Y$. 
The maximum number of outcomes is 604 at age 88, whereas the minimum
is 7 at age 108. We resort to the strategy described in Section
\ref{sec::missingIndiv} and use the scaled loss in \eqref{eq:gnew} to
compensate for the varying sample sizes.  

\subsection{Results}

Out of the 1050 coefficients associated with the predictors described
above, 148 are estimated to be nonzero for at least one time point in
the 34 years age range. More precisely, for at least one age, 57
coefficients are nonzero in the ``dementia vs normal'' equation of the
predictive multinomial logit model, and 124 are nonzero in the ``death
vs normal'' equation. 

\begin{figure}[!ht]
\centering
\includegraphics[width=\textwidth]{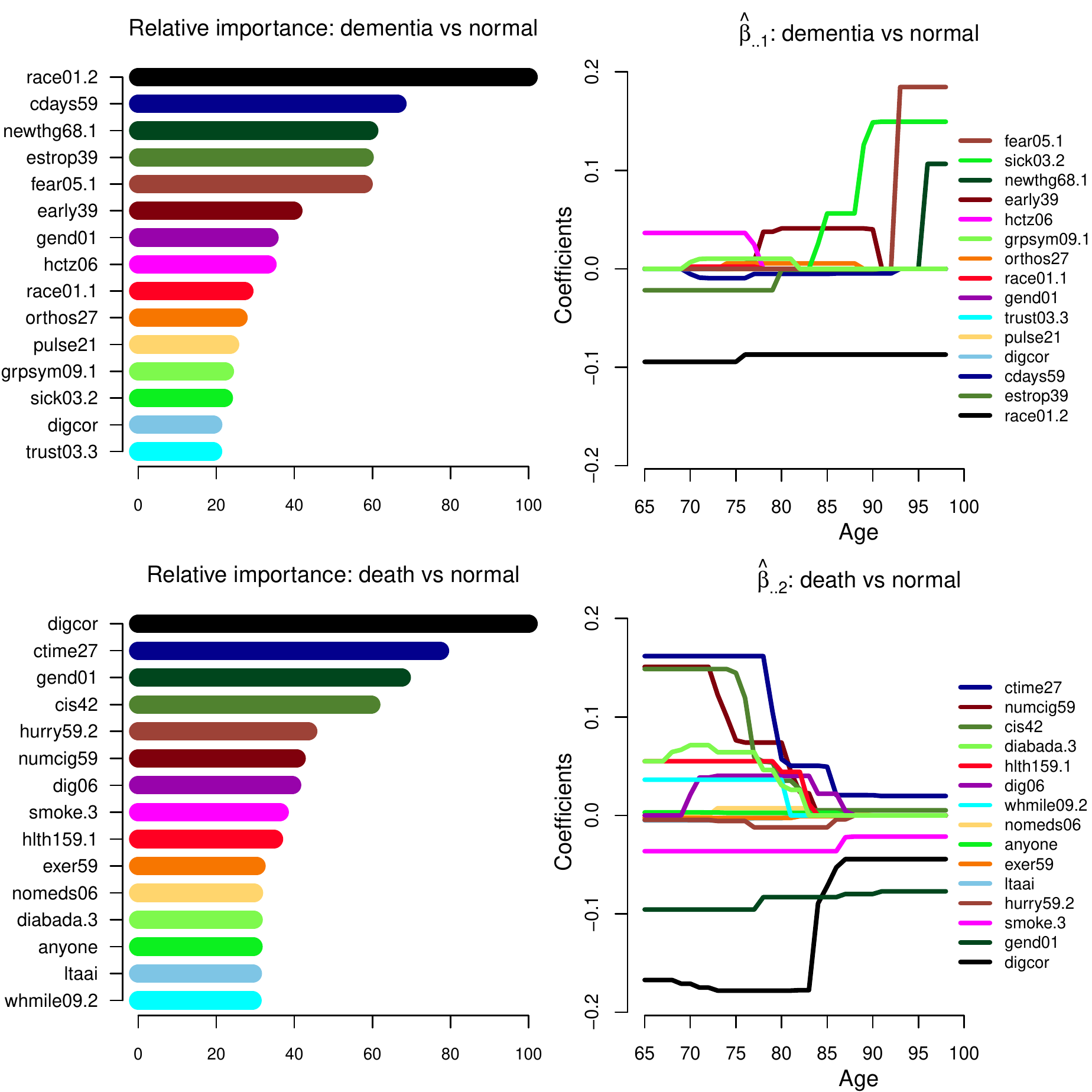}
\caption{CHS-CS data analysis. Left: relative importance plots for the
  15 most important variables in the ``dementia vs normal'' and
  ``death vs normal'' equations of the multinomial logit model. Right:
  corresponding estimated coefficients. The order of the legends
  follow the order of the maximum/minimum values of the estimated
  coefficient trajectories. Note that some coefficients are estimated to be 
  very close to 0 and the corresponding trajectories are hidden by other coefficients.} 
\label{fig::results}
\end{figure}

\begin{table}[!ht]
\centering
\begin{tabular}{| c | l |}
\multicolumn{2}{c}{Dementia vs normal} \\ \hline
\textbf{Variable} & \textbf{Meaning (and coding for categorical
  variables, before scaling)}  \\  \hline 
race01.2 & Race:  "White" 1, else 0 \\ \hline
cdays59 & Taken vitamin C in the last 2 weeks? (number of days)  \\ \hline
newthg68.1 & How is the person at learning new things wrt 10 yrs ago? "A bit worse" 1, else 0 \\ \hline
estrop39 & If you not currently taking estrogen, have you taken in the past? "Yes" 1, "No" 0 \\ \hline
fear05.1 & How often felt fearful during last week? "Most of the time" 1, else 0 \\ \hline
early39 & Do you usually wake up far too early? "Yes" 1, "No" 0 \\ \hline
gend01 & Gender: "Female" 1, "Male" 0 \\ \hline 
hctz06 & Medication: thiazide diuretics w/o K-sparing. "Yes" 1, "No" 0 \\ \hline 
race01.1 & Race:  "Other (no white, no black)" 1, else 0 \\ \hline
orthos27 & Do you use a lower extremity orthosis? "Yes" 1, "No" 0\\ \hline
pulse21 & 60 second heart rate \\ \hline
grpsym09.1 & What causes difficulty in gripping? "Pain in arm/hand" 1, else 0 \\ \hline 
sick03.2 & If sick, could easily find someone to help? "Probably False" 1, else 0  \\ \hline
digcor & Digit-symbol substitution task: number of symbols correctly coded \\ \hline
trust03.3 & There is at at least one person whose advice you really trust. "Probably true" 1, else 0 \\ \hline  
\multicolumn{2}{c}{} \\
\multicolumn{2}{c}{Death vs normal} \\ \hline
\textbf{Variable} & \textbf{Meaning (and coding for categorical
  variables, before scaling)}  \\  \hline 
digcor & Digit-symbol substitution task: number of symbols correctly coded \\ \hline
ctime27 & Repeated chair stands: number of seconds \\ \hline
gend01 & Gender: "Female" 1, "Male" 0 \\ \hline 
cis42 & Cardiac injury score  \\ \hline 
hurry59.2 & Ever had pain in chest when walking uphill/hurry? "No" 1, else 0 \\ \hline
numcig59 & Number of cigarettes smoked per day  \\ \hline
dig06 & Digitalis medicines prescripted?  "Yes" 1, "No" 0  \\ \hline
smoke.3 & Current smoke status: "Never smoked" 1, else 0 \\ \hline
hlth159.1 & Would you say, in general, your health is.. ? "Fair" 1, else 0  \\ \hline
exer59 & If gained/lost weight, was exercise a major factor? "Yes" 1, "No" 0\\ \hline
nomeds06 & Number of medications taken\\ \hline
diabada.3 & ADA diabetic status? "New diabetes" 1, else 0\\ \hline
anyone & Does anyone living with you smoke cigarettes regularly? "Yes" 1, "No" 0\\ \hline
ltaai & Blood pressure variable: left ankle-arm index \\ \hline
whmile09.2 & Do you have difficulty walking one-half a mile? "Yes" 1, else 0 \\ \hline
\end{tabular}
\caption{The 15 most important variables in the two separate equations
  of the multinomial logit model.} 
\label{tab::variables}
\end{table}

Figure \ref{fig::results} shows the 15 most important variables in the
34 years age range, separately for the two equations. The measure of
importance is described in detail in Section \ref{sec::stability} and
is, in fact, a measure of stability of the estimated coefficients,
across 4 subsets of the data (the 4 training sets used
in cross-validation). The plots on the left show the relative
importance of the 15 variables with respect to the most important one,
whose importance was scaled to be 100. The plots on the right show,
separately for the two equations, the longitudinal estimated
coefficients for the 15 most important variables, using the data and
algorithm specification described above. The meaning of these
predictors and the coding used for the categorical variables are
reported in Table \ref{tab::variables}. 
The nonzero coefficients that are not displayed in Figure
\ref{fig::results} are less important (according to our measure of
stability) and, for the vast majority, their absolute values are less
than 0.1. 

We now proceed to interpret the results, keeping in mind that,
ultimately, we are estimating the coefficients of a multinomial logit
model and that the outcome variable is recorded 10 years in the future
with respect to the predictors. 
For example, an increase in the value of a predictor with positive
estimated coefficient in the top right plot of Figure
\ref{fig::results} is associated with an increase of the (10 years
future) odds of dementia with respect to a normal cognitive status.  
In what follows, to facilitate the exposition of results, our
statements are less formal. 

Inspecting the ``dementia vs normal'' plot we see that, 
in general, being Caucasian ({\tt race01.2}) is associated with a
decrease in the odds of dementia, while, after the age of 85, fear
({\tt fear05.1}), lack of available caretakers ({\tt sick03.2}), and
deterioration of learning skills ({\tt newthg68.1}) increase the odds
of dementia. Variables {\tt hctz06} (a particular diuretic) and {\tt
  early39} (early wake-ups) have positive coefficients for the age
ranges $65, \ldots 78$ and $77, \ldots 91$, respectively, and hence,
if active, they account for an increase of the risk of dementia.   
The ``death vs normal" plot reveals the importance of several
variables in the age range $65, \ldots 85$: longer time to rise from sitting in a chair
({\tt ctime27}), more cigarettes ({\tt numcig59}), higher
cardiac injury score ({\tt cis42}) are associated with an increase of
the odds of death. Other variables in the same age range, with analogous
interpretations, but lower importance, are {\tt diabada.3} ("new
diabetes"' diagnosis), {\tt hlth159.1} ("fair" health status), {\tt
  dig06} (use of Digitalis), {\tt whmile09.2} (difficulty in
walking).  
By contrast, in the same age range, good performance on the
digit-symbol substitution task ({\tt digcor}) accounts for a decrease
in the odds of death. Finally, regardless of the age, being a
non-smoker ({\tt smoke.3}) or being a woman ({\tt gend01}) decrease
the odds of death. 

\begin{figure}[!ht]
\centering
\includegraphics[width=\textwidth]{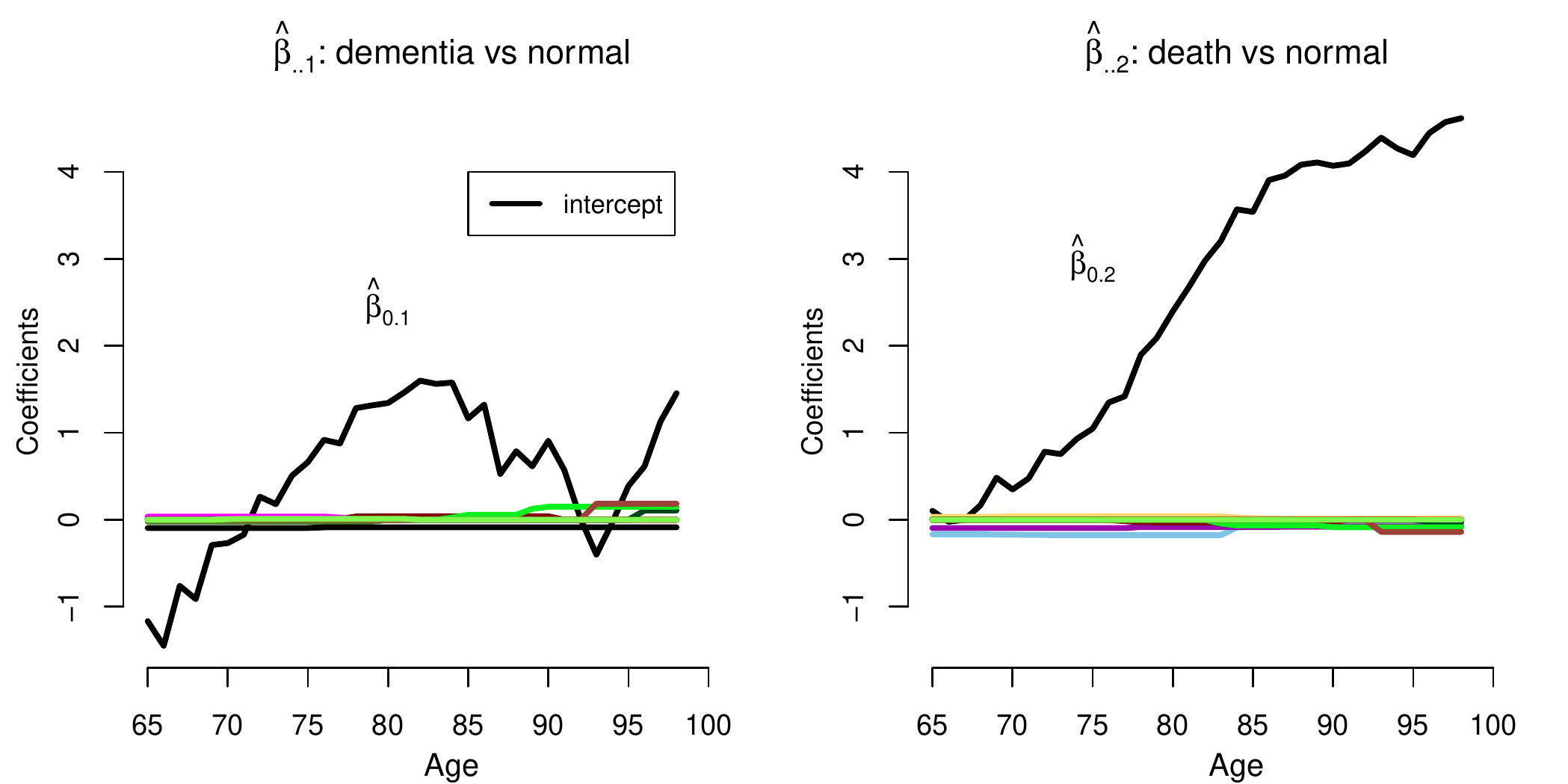}
\caption{CHS-CS data analysis. Estimated intercept coefficients in the
  two separate equations of the multinomial logit model.} 
\label{fig::intercepts}
\end{figure}

Figure \ref{fig::intercepts} shows the intercept coefficients 
\smash{$\hat \beta_{0 \cdot 1}$} and \smash{$\hat \beta_{0 \cdot 2}$},
which, we recall, are not penalized in the log likelihood criterion in
\eqref{eq::fusedmodel}.  The intercepts account for time-varying
risk that is not explained by the predictors.  In particular, the
coefficients $\hat \beta_{0 \cdot 2}$ increases over time, suggesting
that an increasing amount of risk of death can be attributed to a
subject's age alone, independent of the predictor measurements. 

\subsection{Discussion}
% We note here
% that the matrix $Y$ contains more instances of ``death'' (7093) than 
% ``MCI/dementia'' (2235) and ``normal'' (1176). This is expected, since
% most of the individuals in the study eventually died, but not everyone
% was diagnosed with dementia or MCI. 
The results of the proposed multinomial fused lasso methodology
applied to the CHS data are broadly consistent with what is known
about risk and protective factors for dementia in the elderly
\citep{lopez2013patterns}. Race, gender, vascular and heart disease,
lack of available caregivers, and deterioration of learning and memory
are all associated with an increased risk of dementia. 
The results, however, provide critical new insights into the natural
progression of MCI/dementia. First, the relative importance of the
risk factors changes over time. As shown in Figure \ref{fig::results},
with the exception of race, risk factors for dementia become more
relevant after the age of 85. This is critical, as there is increasing
evidence \citep{kuller2011does} for a change in the risk profile for
the expression of clinical dementia among the oldest-old. 
Second, the independent prediction of death, and the associated
risk/protection factors, highlight the close connection between risk
of death and risk of dementia. That is, performance on a simple, timed
test of psychomotor speed (digit symbol substitution task) is a very 
powerful predictor of death within 10 years, as is a measure of
physical strength/frailty (time to arise from a chair). Other
variables, including gender, diabetes, walking and exercise, are all
predictors of death, but are known, from other analyses in the CHS and
other studies, to be linked to the risk of dementia. The importance of
these risk/protective factors for death is attenuated (with the
exception of gender) after age 85, likely reflecting survivor bias. 
Taken together, these results add to the growing body of evidence 
of the critical importance of accounting for
mortality in the analysis of risk for dementia, especially among the
oldest old \citep{kuller2011does}.
%The fact that the relative importance of the risk/protective factors
%for both dementia/MCI and death change as a function of age, suggests
%that \todo{change this} we may need to reconsider some of our
%formulations regarding the natural history of dementia, which assume
%equivalent importance of risk/protection across the age range. 

For our analysis we chose a 10 year time window for risk
prediction. Among individuals of age 65-75, who are cognitively
normal, this may be a scientifically and clinically reasonable time
window to use. However, had we similar data from individuals as young
as 45-50 years old, then we might wish to choose time windows of 20
years or longer. In the present case, it could be argued that a
shorter time window might be more scientifically and clinically
relevant among individuals over the age of 80 years, as survival times
of 10 years become increasingly less likely in the oldest-old.

\section{Measures of stability}
\label{sec::stability}

Examining the stability of variables in a fitted model, subject to
small perturbations of the data set, is one way to assess variable
importance. Applications of stability, in this spirit, have recently
gained popularity in the literature, across a variety of settings such
as clustering (e.g., \citet{stabcluster}), regression (e.g.,
\citet{stabselect}), and graphical models (e.g., \citet{stars}).  Here
we propose a very simple stability-based measure of variable
importance, based on the definition of variable importance for trees
and additive tree expansions \citep{cart,esl}.  We fit the multinomial
fused lasso estimate \eqref{eq::fusedmodel} on the data set
$X_{i\cdot\cdot}$, $Y_{i\cdot}$, for $i=i_1,\ldots i_m$, a subsample
of the total individuals $1,\ldots n$, and repeat this process $R$
times. Let \smash{$\hbeta^{(r)}$} denote the coefficients from the
$r$th subsampled data set, for $r=1,\ldots R$.  Then we define the
importance of variable $j$ for class $k$ as 
\begin{equation}
\label{eq::ij}
I_{jk} =  \frac{1}{RT} \sum_{r=1}^R \sum_{t=1}^T
|\hbeta_{jtk}^{(r)}|, 
\end{equation}
for each $j=1,\ldots p$ and $k=1, \ldots K-1$, which is the average
absolute magnitude of the coefficients for the $j$th variable and
$k$th class, across all timepoints, and subsampled data sets.
Therefore, a larger value of $I_{jk}$ indicates a higher variable
importance, as measured by stability (not only across subsampled data
sets $r$, but actually across timepoints $t$, as well). Relative
importances can be computed by scaling the highest variable importance
to be 100, and adjusting the other values accordingly; for simplicity
we typically consider relative variable importances in favor of
absolute ones, because the original scale has no real meaning. 

There is some subtlety in the role of the tuning parameters
$\lambda_1,\lambda_2$ used to fit the coefficients
\smash{$\hbeta^{(r)}$} on each subsampled data set $r=1,\ldots
R$. Note that the importance measure \eqref{eq::ij} reflects the
importance of a variable in the context of a fitting procedure that,
given data samples, produces estimated coefficients.  The simplest
approach would be to consider the fitting procedure defined by the
multinomial fused lasso problem \eqref{eq::fusedmodel} at a fixed pair
of tuning parameter values $\lambda_1,\lambda_2$. But in practice, it
is seldom true that appropriate tuning parameter values are known
ahead of time, and one typically employs a method like
cross-validation to select parameter values (see Section
\ref{sec::selection} for a discussion of cross-validation and other
model selection methods).  Hence in this case, to determine variable
importances in the final coefficient estimates, we would take care to
define our fitting procedure in \eqref{eq::ij} 
%under consideration 
to be the one that, given data samples, performs cross-validation on
these data samples to determine the best choice of
$\lambda_1,\lambda_2$, and then uses this choice to fit coefficient
estimates.  In other words, for each subsampled data set $r=1,\ldots
R$ in \eqref{eq::ij}, we would perform cross-validation to determine
tuning parameter values and then compute \smash{$\hbeta^{(r)}$} as the
multinomial fused lasso solution at these chosen parameter
values. This is more computationally demanding, but it is a more
accurate reflection of variable importance in the final model output
by the multinomial fused lasso under cross-validation for model
selection. 

The relative variable importances for the CHS-CS data example from
Section \ref{sec::CHS} are displayed in Figure \ref{fig::results},
alongside the plots of estimated coefficients. Here we drew 4
subsampled data sets, each one containing 75\% of the total number of
individuals. The tuning parameter values have been selected by
cross-validation. The variable importances were defined to incorporate
this selection step into the fitting procedure, as explained
above. We observe that the variables with high positive or negative
coefficients for most ages in the plotted trajectories typically also
have among the highest relative importances.  Another interesting
observation concerns categorical predictors, which (recall) have been
converted into binary predictors over multiple levels: often only some
levels of a categorical predictor are active in the plotted
trajectories. 

%NOT TRUE: but in such cases, most or all levels of the variable are
%assigned comparable relative importances.   

%%%%%%%%%%%%%%%
\section{Model selection} 
%%%%%%%%%%%%%
\label{sec::selection}

The selection of tuning parameters $\lambda_1,\lambda_2$ is clearly an 
important issue that we have not yet covered.  In this section, we
discuss various methods for automatic tuning parameter selection in
the multinomial fused lasso model \eqref{eq::fusedmodel}, and apply
them to a subset of the CHS-CS study data of Section \ref{sec::CHS},
with 140 predictors and 600 randomly selected individuals, 
as an illustration.  In particular, we consider the following methods
for model selection: cross-validation, cross-validation under the
one-standard-error rule, AIC, BIC, and finally AIC and BIC using
misclassification loss (in place of the usual negative log
likelihood).  Note that cross-validation in our longitudinal setting
is performed by dividing the individuals $1,\ldots n$ into folds, and,
per its typical usage, selecting the tuning parameter pair
$\lambda_1,\lambda_2$ (over, say, a grid of possible values) that
minimizes the cross-validation misclassification loss.   The
one-standard-error rule, on the other hand, picks the simplest
estimate that achieves a cross-validation misclassification loss
within one standard error of the minimum.  Here ``simplest'' is
interpreted to mean the estimate with the fewest number of nonzero
component blocks.  AIC and BIC scores are computed for a candidate
$\lambda_1,\lambda_2$ pair by 
\begin{align*} 
\mathrm{AIC}(\lambda_1,\lambda_2) &= 2 \cdot 
\mathrm{loss}\big((\hbeta_0,\hbeta)_{\lambda_1,\lambda_2}\big) \,+\, 
2 \cdot \mathrm{df}\big((\hbeta_0,\hbeta)_{\lambda_1,\lambda_2}\big), \\ 
\mathrm{BIC}(\lambda_1,\lambda_2) &= 2 \cdot  
\mathrm{loss}\big((\hbeta_0,\hbeta)_{\lambda_1,\lambda_2}\big) \,+\,
\log{N_\mathrm{tot}} \cdot  
\mathrm{df}\big((\hbeta_0,\hbeta)_{\lambda_1,\lambda_2}\big), 
\end{align*} 
and in each case, the tuning parameter pair is chosen (again, say,
over a grid of possible values) to minimize the score. 
In the above, \smash{$(\hbeta_0,\hbeta)_{\lambda_1,\lambda_2}$}
denotes the multinomial fused lasso estimate \eqref{eq::fusedmodel} at
the tuning parameter pair $\lambda_1,\lambda_2$, and $N_\mathrm{tot}$
denotes the total number of observations in the longitudinal study,
$N_\mathrm{tot}=nT$ (or \smash{$N_\mathrm{tot}=\sum_{t=1}^T n_t$} in
the missing data setting).  Also,
\smash{$\mathrm{df}((\hbeta_0,\hbeta)_{\lambda_1,\lambda_2})$} denotes
the degrees of freedom of the estimate
\smash{$(\hbeta_0,\hbeta)_{\lambda_1,\lambda_2}$}, and we employ the
approximation 
\begin{equation*}
\mathrm{df}\big((\hbeta_0,\hbeta)_{\lambda_1,\lambda_2}\big) \approx 
\text{\# of nonzero blocks in $(\hbeta_0,\hbeta)_{\lambda_1,\lambda_2}$},
\end{equation*}
borrowing from known results in the Gaussian likelihood case
\citep{genlasso,lassodf2}.  Finally,  
\smash{$\mathrm{loss}((\hbeta_0,\hbeta)_{\lambda_1,\lambda_2})$}
denotes a loss function considered for the estimate, which we take
either to be the negative multinomial log likelihood
\smash{$-\ell((\hbeta_0,\hbeta)_{\lambda_1,\lambda_2})$}, as is
typical in  AIC and BIC \citep{esl}, or the misclassification loss, to
put it on closer footing to cross-validation.  Note that both loss
functions are computed in-sample, i.e., over the training samples, and
hence AIC and BIC are computationally much cheaper than
cross-validation.   

\begin{figure}[!ht]
\centering
\includegraphics[width=\textwidth]{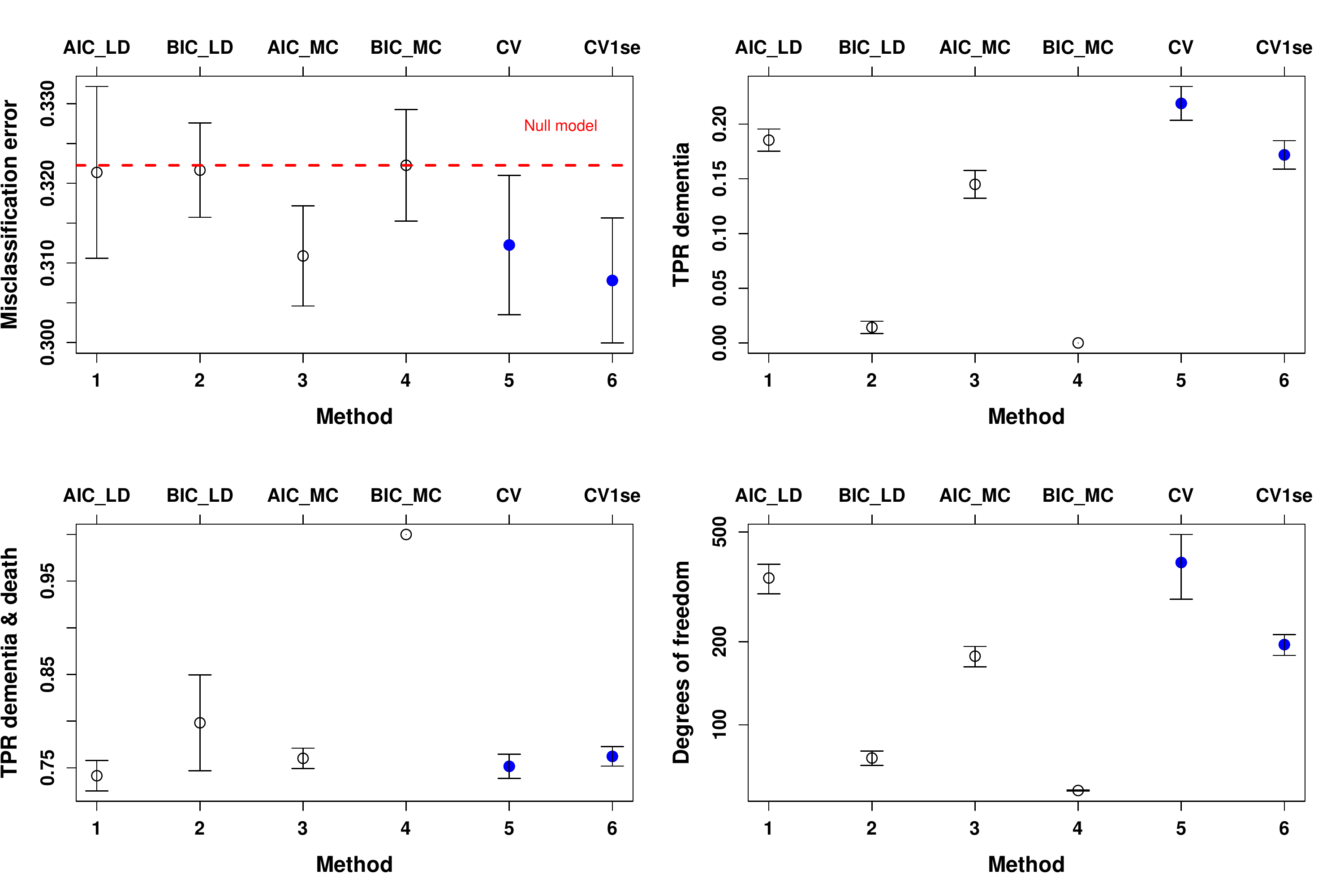}
\caption{
Comparison of different methods for selection of tuning parameters
$\lambda_1,\lambda_2$ on the CHS-CS data set.  The x-axis in each plot
parametrizes the 6 different methods considered, which are, from left
to right: AIC and BIC under negative log likelihood loss, AIC and BIC
under misclassification loss, cross-validation, and cross-validation
with the one-standard-error rule.  The upper left plot shows 
(out-of-sample) misclassification rate associated with the estimates
selected by each method, averaged over 5 iterations.  The segments
denote $\pm 1$ standard errors around the mean.  The red dotted line
is the average misclassification rate associated with the naive
estimator that predicts all individuals as dead (the majority class).
The upper right and bottom left plots show different measures of
evaluation (again, computed out-of-sample): the true positive rate in
identifying the dementia class, respectively, the true positive rate
in identifying the dementia and death classes combined.  Finally, the
bottom right plot shows degrees of freedom (number of nonzero blocks)
of the estimates selected by each method.} 
\label{fig::selection} 
\end{figure}

We compare these model selection methods on the subset of the CHS-CS data set.  The
individuals are randomly split into 5 folds.  We use 4/5 of the data
set to perform model selection and subsequent model fitting with the 6
techniques described above: cross-validation, cross-validation with
the one-standard-error rule, and AIC and BIC under negative log
likelihood and misclassification losses.  To be perfectly clear, the
model selection techniques work entirely within this given 4/5 of the
data set, so that, e.g., cross-validation further divides this data
set into folds.  In fact, we used 4-fold cross-validation to make this
division simplest.  The remaining 1/5 of the data set is then used for
evaluation of the estimates coming from each of the 6 methods, and
this entire process is repeated, leaving out each fold in turn as the
evaluation set.  We record several measures on each evaluation set:
the misclassification rate, true positive rate in identifying the
dementia class, true positive positive rate in identifying the
dementia and death classes combined, and degrees of freedom (number of
nonzero blocks in the estimate).  Figure \ref{fig::selection} displays
the mean and standard errors of these 4 measures, for each of the 6
model selection methods.   

%\todo{Just for a check, not that we need to put this in: how were the
%standard errors computed?  Are they standard deviations, or standard
%deviations / sqrt(5)?}{\color{blue}They are standard
%deviations/sqrt(5): SA} 

Cross-validation and cross-validation with the
one-standard-error rule both seem to represent a favorable balance 
between the different evaluation measures.  The cross-validation
methods provide a misclassification rate significantly better than
that of the null model, which predicts according to the majority class
(death), they yield two of the three highest true positive rates in
identifying the dementia class, and perform well in terms of
identifying the dementa and death classes combined (as do all methods:
note that all true positive rates here are about 0.75 or higher).
We ended up settling cross-validation under the usual rule, rather
than the one-standard-error rule, because the
former achieves the highest true positive rate in identifying the
dementia class, which was our primary concern in the CHS-CS data
analysis.  By design, cross-validation with the one-standard-error
rule delivers a simpler estimate in terms of degrees of freedom (196
for the one-standard-error rule versus 388 for the usual rule) though
both cross-validation models are highly regularized in absolute
terms (e.g., the fully saturated model would have thousands of nonzero  
blocks).  

% However, cross-validation with the one standard error rule
% achieves an average degrees of freedom of 196, higher than that of
% most of the information criteria estimates (with an exception of AIC
% with logdeviance loss function), which are between 58 and 177, but
% much smaller than that of the usual cross-validation estimate, which
% is around 388. While cross-validation with the one standard error rule
% provides relatively simple model (in terms of the degrees of freedom),
% the model selected by this method has less temporal information than
% the one selected by cross-validation. Empirically we observed that the
% estimates of the coefficients under crossvalidation with the one
% standard error rule for many of the selected predictors are constant
% over time. On the CHS-CS data set, we seek for a result that is sparse
% and interpretable, yet biologically informative and interesting. The
% model selected by cross-validation seems to satisfy both
% criteria. Hence, we proceeded by using this method to select tuning
% parameters on the entire data set by 4-fold cross-validation, and the
% results were shown in Section \ref{sec::CHS}. 

\section{Discussion and future work} 
\label{sec::discussion}

In this work, we proposed a multinomial model for
high-dimensional longitudinal classification tasks.  Our proposal
operates under the assumption that a sparse number of predictors
contribute more or less persistent effects across time.  The
multinomial model is fit under lasso and fused lasso regularization,  
which address the assumptions of sparsity and persistence,
respectively, and lead to piecewise constant estimated coefficient 
profiles.  We described a highly efficient computational algorithm for
this model based on proximal gradient descent, demonstrated the
applicability of this model on an Alzheimer's data set taken from the
CHS-CS, and discussed practically important issues such stability
measures for the estimates and tuning parameter selection. 

A number of extensions of the basic model are well within reach.
For example, placing a group lasso penalty on the coefficients
associated with each level of a binary expansion for a categorical
variable may be useful for encouraging sparsity in a group sense
(i.e., over all levels of a categorical variable at once).  As another
example, more complex trends than piecewise constant ones may be fit
by replacing the fused lasso penalty with a trend filtering penalty 
\citep{l1tf,trendfilter}, which would lead to piecewise polynomial
trends of any chosen order $k$.  The appropriateness of such a penalty
would depend on the scientific application; the use of a fused lasso
penalty assumes that the effect of a given variable is mostly constant
across time, with possible change points; the use of a quadratic trend
filtering penalty (polynomial order $k=2$) allows the effect to vary
more smoothly across time.  

More difficult and open-ended extensions concern statistical inference
for the fitted longitudinal classification models.  For example, the
construction of confidence intervals (or bands) for selected
coefficients (or coefficient profiles) would be an extremely useful
tool for the practitioner, and would offer more concrete and
rigorous interpretations than the stability measures described in
Section \ref{sec::stability}.  Unfortunately, this is quite a
difficult problem, even for simpler regularization schemes (such as
a pure lasso penalty) and simpler observation models (such as linear 
regression).  But recent inferential developments for related
high-dimensional estimation tasks
\citep{zhangconf,montahypo2,vdgsignif,LTTT2013,exactlasso,exactlars} 
shed a positive light on this future endeavor.

%\section*{Acknowledgements}
%We would like to sincerely thank XXXX

\bibliographystyle{plainnat}
\bibliography{paper,ryantibs}

\end{document}